\documentclass[]{raa}            
\usepackage{graphicx,times}             

\begin{document}

   \title{The simplest way to get a cluster's parameters in the {\em Gaia} era (Dolidze 41)
}
   \volnopage{Vol.0 (2018) No.0, 000--000}      
   \setcounter{page}{1}          

   \author{Tadross, A. L.
        }
   \institute{National Research Institute of Astronomy \& Geophysics, 11421-Helwan, Cairo, Egypt. \\ {\it altadross@yahoo.com}\\
           }

   \date{Received~~2009 month day; accepted~~2009~~month day}

\abstract{The astro-photometric parameters of the open star cluster Dolidze 41, which located in the constellation of Cygnus, have been investigated using the {\em Gaia-ESO} DR2 large Survey merging with the near Infrared Two Micron All Sky Survey {\em 2MASS} database. The radial density distribution (limited, core and tidal radii), color-magnitude diagrams, the galactocentric coordinates, distances, color excess, and age of Dolidze 41 are presented. Thanks to {\em Gaia} DR2 astrometry, which help us to define the membership of the cluster stars easily. The luminosity \& mass functions, the entire luminosity \& mass, and the repose time of the cluster have been estimated as well.
}

   \authorrunning{Tadross, A. L.}            
   \titlerunning{The simplest way to get a cluster's parameters in the {\em Gaia} era (Dolidze 41)}  

  \maketitle
\section{Introduction}

\begin{figure}[h]
\centering
\includegraphics[width=10cm, height=10cm]{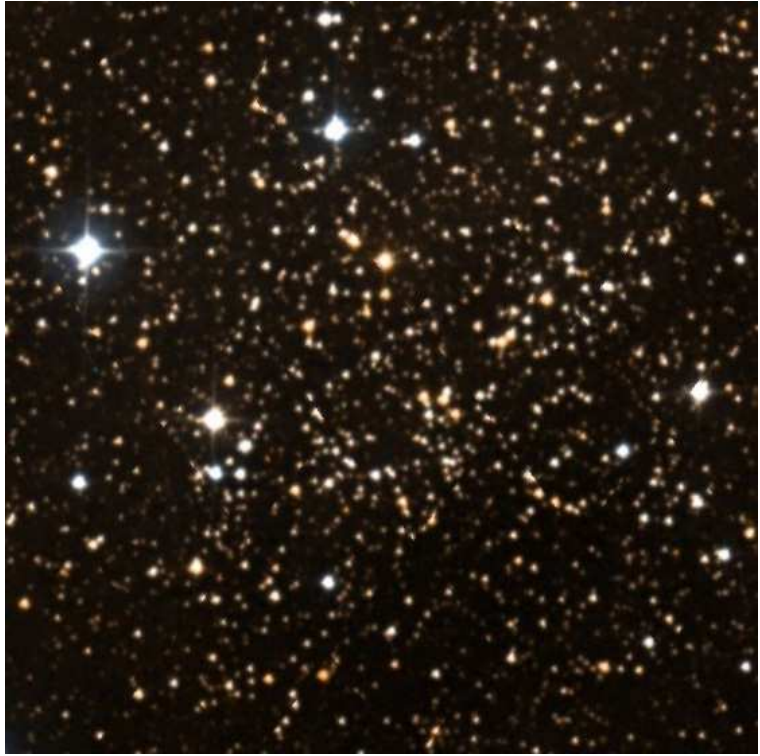}
\caption{The image of Dolidze 41 as taken from {\em ALADIN}. $\alpha=20^{h}\ 18^{m}\ 49^{s},\ \delta=+37^{\circ}\ 45^{'}\ 00^{''}, \ell= 75.707^{\circ} ~\& \ b= 0.9925^{\circ}$. North is up, East to the left.}
 \label{Fig-1}
 \end{figure}

One of the foremost necessary constituents in studying the structure and evolution of the Milky Way system are star clusters. Known stellar clusters comprise about 160 globular clusters and about 3000 open clusters \cite{2013yCat..35580053K}. If we extrapolate the solar vicinity to the whole disc, we may reach about 100,000 open clusters \cite{2018IAUS..330..119B}. Many of such objects are newly discovered and need photometric investigations, as well as confirmation of their physical nature. We believe that most of them will be detected and investigated in the {\em Gaia} DR2 era.
The name {\em GAIA} was originally derived as an acronym for Global Astrometric Interferometer for Astrophysics. This reflected the optical technique of interferometry that was originally planned for use on the spacecraft. While the working method evolved during studies and the acronym is no longer applicable; however, the name {\em GAIA} remained to provide continuity with the project. It is the backbone in the science program of the European Space Operations (ESO), which launched on 19 December 2013, and situated 1.5 million km from Earth. The spacecraft {\em GAIA} monitored each of its target objects about 70 times to a magnitude of G = 20 over a period of five years to study the precise position and motion of each one. The {\em Gaia} DR2 was released on 25 April 2018 for 1.7 billion exquisite precision sources in astrometric five-parameter solutions of coordinates, proper motions in right ascension and declination, and parallaxes ($\alpha$, $\delta$, $\mu_\alpha \cos{\delta}$, $\mu_\delta$, $\pi$). In addition, magnitudes in three photometric filters ($G, G_{BP}, G_{RP}$) for more than 1.3 billion sources \cite{2018arXiv180409365G}. {\em Gaia} Archive is available through the web page {\em (http://www.cosmos.esa.int/gaia)}.
\\ \\
With the help of the Virtual Observatory tool {\em TOPCAT} and {\em ALADIN} we could use the cross-matched data of {\em Gaia} DR2 and the {\em 2MASS} surveys to compile a useful photometrical data to investigate Dolidze 41 \cite{2000A&AS..143...33B, 2005ASPC..347...29T}. The current work could be a part of our continued series whose goal is to get the most astrophysical properties of newly, antecedently unstudied and/or poorly studied open clusters utilizing the foremost newly databases \cite{2008NewA...13..370T, 2008MNRAS.389..285T, 2009NewA...14..200T, 2009Ap&SS.323..383T, 2011JKAS...44....1T}. The most important aspect of using {\em Gaia} DR2 with the {\em 2MASS} surveys lies in the positions, parallax and proper motions for the cluster' stars, which makes the member candidates can be easily determined.
\\
\\
J2000.0 coordinates of Dolidze 41 are $\alpha=20^{h}\ 18^{m}\ 49^{s},\ \delta=+37^{\circ}\ 45^{'}\ 00^{''},\ \ell= 75.707^{\circ} ~\& \ b= 0.9925^{\circ}$ in the Cygnus constellation with a roughly angular size of 12 arcmin as shown in Fig.~\ref{Fig-1}. Kazlauskas et al. \cite{2013NewA...19...34K} and Dias et al. \cite{2014A&A...564A..79D} mentioned some few astrometric information about the cluster. No real photometric data has been published for Dolidze 41 to date except the study of Tadross \& Nasser \cite{2010arXiv1011.2934T}, hereafter TN (2010). In the present work, we re-study Dolidze 41 in the light of {\em Gaia} DR2 database. The near-infrared {\em JHK} photometry of the 2MASS catalog of \cite{2006AJ....131.1163S} is also used to estimate and confirm the most main properties of the cluster. \\

This paper is arranged as follows: The target data is presented in Section 2. The radial density profile is described in Section 3. Color-Magnitude Diagrams are presented in Section 4. Luminosity - mass functions and the dynamical status of the cluster are discussed in Section 5. Finally, the conclusion of the present study is given in Section 6.
\section{Target Data}

\begin{figure}[h]
\centering
\includegraphics[width=16cm]{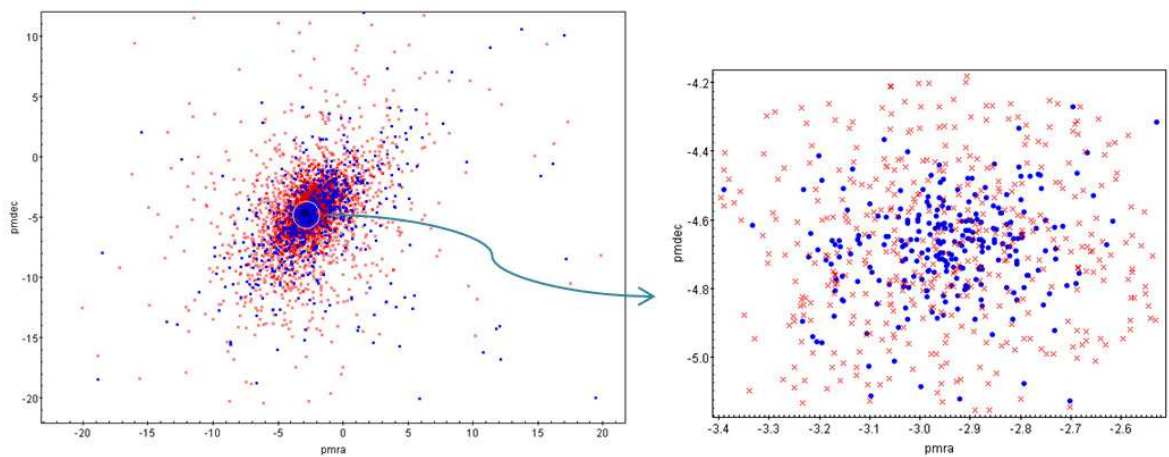}
\caption{The left hand panel shows the proper motion diagram for Dolidze 41. The red crosses refer to the concentrated field with halo-like background field, while the blue dots show the stars with proper motion errors less than 10\%. The white circle refers to the studied cluster's area, in which 480 co-moving stars are located as shown in the zooming view of the right hand panel.}
 \label{Fig-2}
\end{figure}

\begin{figure}[h]
\centering
\includegraphics[width=12cm, height=7cm]{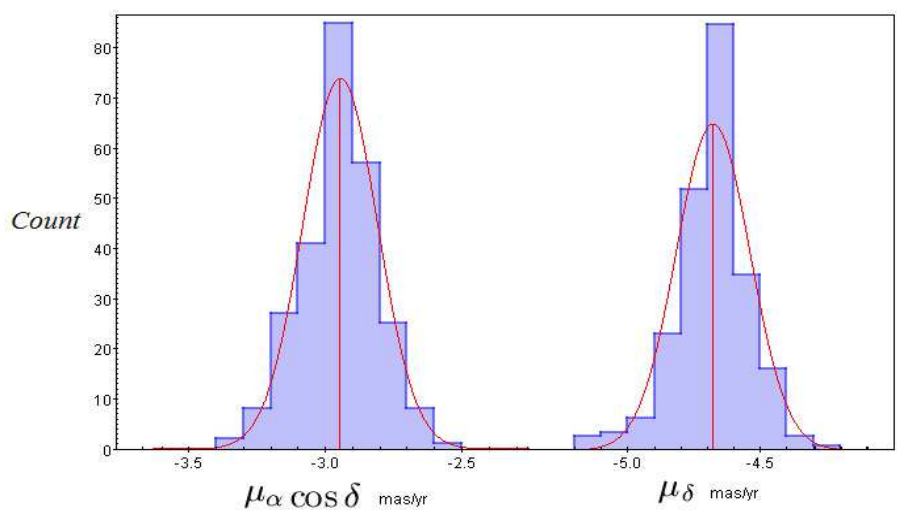}
\caption{The mean values of $\mu_\alpha \cos{\delta}$, and $\mu_\delta$ for the cluster member candidates, which are found to be -2.95 $\pm$ 0.14 mas/yr and -4.68 $\pm$ 0.14 mas/yr respectively.}
 \label{Fig-3}
\end{figure}

\begin{figure}[h]
\centering
\includegraphics[width=15cm, height=7cm]{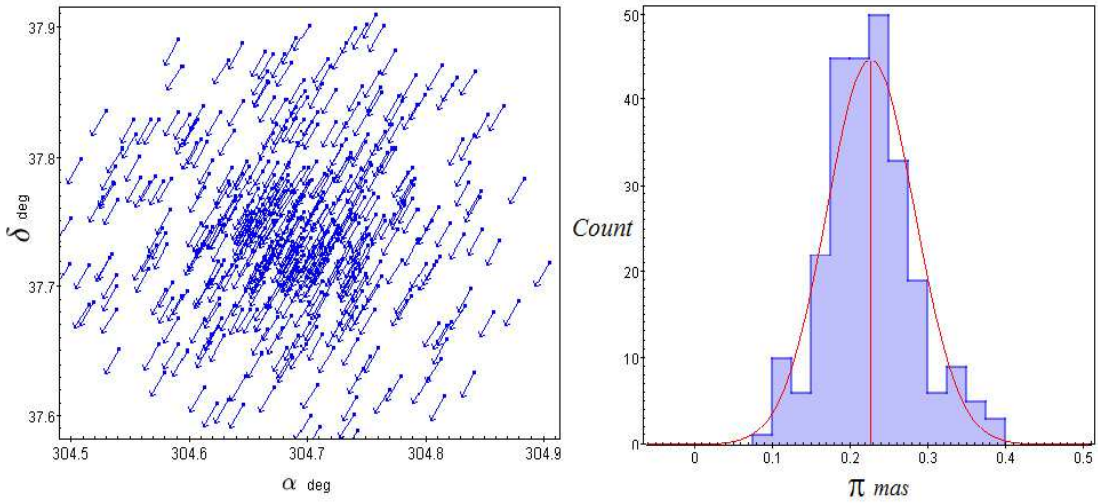}
\caption{The left hand panel shows the magnified velocities of the cluster member candidates, i.e. the co-moving stars within the cluster area. They are located in the rang of the parallax of 0.1\,mas $< \pi <$ 0.4\,mas, which presents in the right hand panel; the mean parallax = 0.23 $\pm$ 0.06 mas.}
 \label{Fig-4}
\end{figure}

Using ({\em TOPCAT}), 12,625 sources were downloaded from {\em Gaia} DR2 database service within a size of 10 arcmin located at the center of the cluster. While, 6,097 near-infrared sources were downloaded from {\em 2MASS} database service \cite{2006AJ....131.1163S} for the same area of the cluster. Both data were cross-matched, getting 5,570 sources, which contains all the datum we need. Of the {\em Gaia} DR2-2MASS cross-matched sources, a subset of 480 stars around the cluster center were selected as co-moving stars (the stars who travel together in the same direction through space) with halo-like background field \cite{2018IAUS..330..119B}, as shown in Fig.~\ref{Fig-2}. For those co-moving stars, the mean values and the standard deviations of $\mu_\alpha \cos{\delta}$, and $\mu_\delta$ were found to be -2.95 $\pm$ 0.14 mas/yr and -4.68 $\pm$ 0.14 mas/yr respectively, as shown in Fig.~\ref{Fig-3}. Stars are considered cluster's member candidates if its 3$\sigma$ parallax error lies within the cluster mean parallax of 0.23 $\pm$ 0.06 mas; located in the ranging of 0.1\,mas $<\pi<$ 0.4\,mas. Those stars were selected to be the cluster's member candidates, which have almost the same speed and direction in the sky with respect to the fore/ background field ones, and at the same time located within the limited size of the cluster, see Fig.~\ref{Fig-4}.

\section{Radial Density Profile}

\begin{figure}[h]
\centering
\includegraphics[width=9cm, height=7cm]{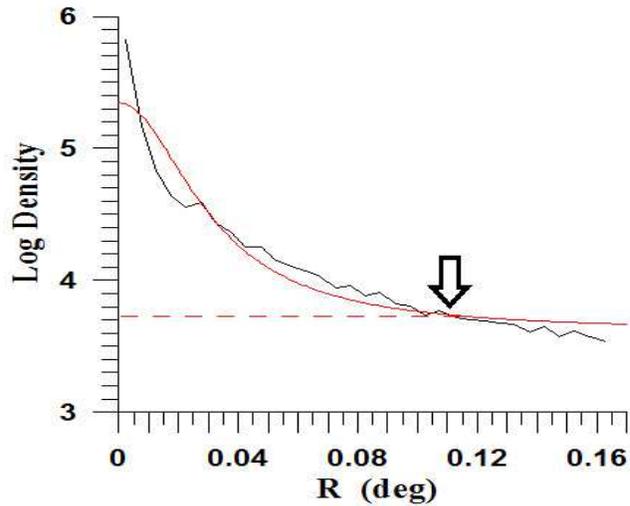}
\caption{The radial density profile of Dolidze 41. The model curve of King \cite{1966AJ.....71...64K}  has been applied. The angular limited radius $R_{lim}$ = 0.11$^{\circ}$ (6.6 arcmin, i.e. $\sim$ 8.8 pc). The core radius $r_{c}$ = 0.03$^{\circ}$ (1.8 arcmin, i.e. $\sim$ 2.4 pc.)}
 \label{Fig-5}
\end{figure}

The radius determination is one of the most important fundamental properties of the cluster. To determine the cluster's limited radius and core radius, the radial surface density of the stars $\rho(r)$ should be achieved firstly. In this context, the stars inside Dolidze 41 area are counted in concentric rings outwards from the cluster center with equal increment radius of 0.1 arcmin. Fig.~\ref{Fig-5} shows the radial density profile of the cluster, where the mean stellar density in each ring is plotted against the corresponding average radius. The cluster limited border is taken at the point, which comprises the entire cluster area and arrives at a sufficient stability with the background field density, i.e. at that point where the cluster stars dissolved in the background field; for more details see \cite{2011JKAS...44....1T}. Applying the empirical King's Model \cite{1966AJ.....71...64K}, the density function $\rho(r)$ is represented as:

\begin{equation}
\rho(r)=f_{bg}+\frac{f_{0}}{1+(r/r_{c})^{2}}
\end{equation}

where $f_{bg}$, $f_{0}$ and $r_{c}$ are background, central densities of the stars in the cluster and the core radius of the cluster respectively. The angular limited radius is found to be 6.6 arcmin ($\sim$ 8.8 pc), while the core radius $r_{c}$ = 1.8 arcmin ($\sim$ 2.4 pc). The concentration parameter of Peterson \& King \cite{1975AJ.....80..427P}, $C$ = log($R_{lim}/R_{c}$) $\approx$ 0.6.

\section{Colour-Magnitude Diagrams}

\begin{figure}[h]
\centering
\includegraphics[width=14cm, height=7cm]{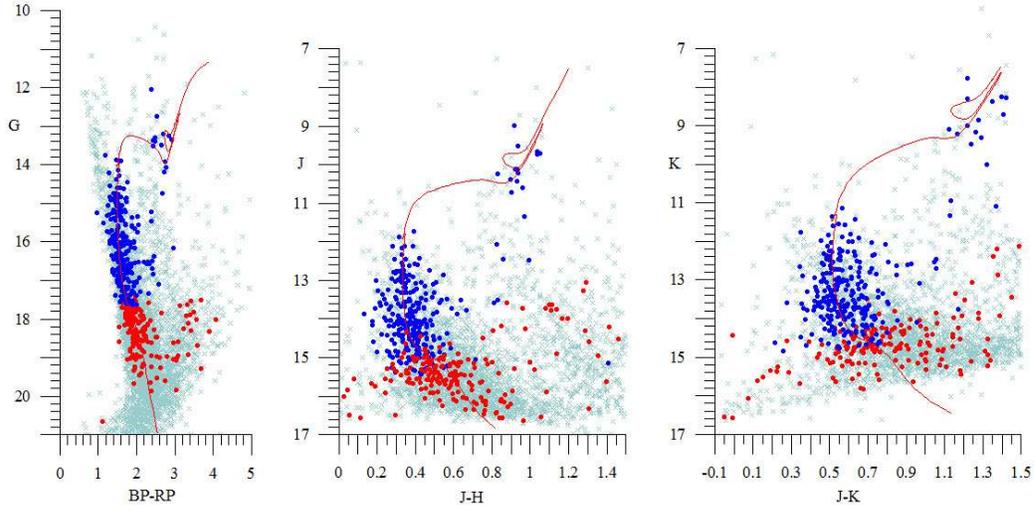}
\caption{The CMDs of the cluster Dolidze 41; the blue dots refer to the co-moving stars in the cluster's area; located within the ranging parallax of the cluster and the proper motion errors less than 10\%. The red dots refer to the stars with higher proper motion errors. The cyan dots refer to the fore/ background field stars around the cluster region. The left hand panel shows the CMD of {\em Gaia} DR2 bandpass G$\sim$(BP-RP), while the middle and right hand panels show the CMDs of {\em 2MASS} bandpass J$\sim$(J-H) \& K$\sim$(J-K). All diagrams fitted to the theoretical solar isochrones of Marigo \cite{{2017ApJ...835...77M}}. The intrinsic distance modulus is found to be 13.3 $\pm$ 0.1 mag, and the color excess values E(J-H) \& E(J-K) are found to be 0.35 \& 0.55 mag respectively. The age of the cluster is found to be 200 $\pm$ 10 Myr.}
 \label{Fig-6}
\end{figure}

The Color-Magnitude Diagrams (CMDs) of the cluster Dolidze 41 have been constructed using the {\em Gaia} DR2 bandpass G$\sim$(BP-RP) as shown in the left hand panel of Fig.~\ref{Fig-6}, while the middle and right hand panels show the CMDs of {\em 2MASS} bandpass J$\sim$(J-H) \& K$\sim$(J-K). Isochrones concerning to {\em Gaia} data were obtained from CMD 3.0 form of different ages and metallicities at $(http://stev.oapd.inaf.it/cgi-bin/cmd\_3.0)$. All diagrams fitted to the solar theoretical isochrones of Marigo et al. \cite{{2017ApJ...835...77M}}. The 2MASS dataset has the benefits of being homogenized all sky and covering near infrared wavelengths $J$~(1.25~$\mu$m), $H$~(1.65~$\mu$m) and $K_{s}$~(2.16~$\mu$m), wherever young clusters can be observed in their nebulous environments. Isochrone fitting was started using some guesses of ages and metallicities ranges for the cluster. However, we used the photometric systems with solar metallicity, because most open cluster studies based on isochrones fitting of solar metallicity $Z$ = 0.0152  \cite{2009A&A...498..877C, 2011SoPh..268..255C} for simplicity and comparing \cite{2018IAUS..330..281Y}. Best fitted isochrones were then used to estimate the parameters of Dolidze 41.
\\
\\
In Fig.~\ref{Fig-6}, the blue dots refer to the co-moving stars in the cluster's area, located within the ranging parallax of the cluster (0.1\,mas $<\pi<$ 0.4\,mas) and proper motion errors $\leq$ 10\%. While the red dots refer to the stars with higher proper motion errors. The cyan dots refer to the fore/ background field stars in the cluster region. The age of the cluster is found to be 200 $\pm$ 10 Myr. The intrinsic distance modulus $(m-M)_o$ = 13.3 $\pm$ 0.1 mag, which corresponds a distance of 4625$\pm$ 210 pc. The $JHK$ photometric color excess E(J-H) and E(J-K) are found to be 0.35 and 0.55 mag respectively, which corresponds to an optical reddening of E(B-V) = 0.54 mag. Correspondingly, the linear diameter, the distance from galactic center $R_g$, the distance from the galactic plane Z$_\odot$, and the distances X$_\odot$ \& Y$_\odot$ from the Sun on the galactic plane are found to be 17.7 pc, 8.5 Kpc, 80 pc, 1140 pc \& 4480 pc respectively; for more details about the calculations, see \cite{2011JKAS...44....1T}.

\section{Luminosity, Mass Functions, and the dynamical status}

The open cluster represents of hundreds of stars having identical ages and compositions but different masses. The luminosity and mass functions (LF \& MF) depended principally on the determination of the membership of the cluster. Here, the cluster's member candidates are the co-moving stars located within the cluster's area, at intervals the ranging of the cluster's parallax, and proper motion errors $\leq$ 10\%. Consequently, a number of 480 stars in the cluster region are counted inside the limit diameter of 17.7 pc.

\begin{figure}[h]
\centering
\includegraphics[width=9cm, height=10cm]{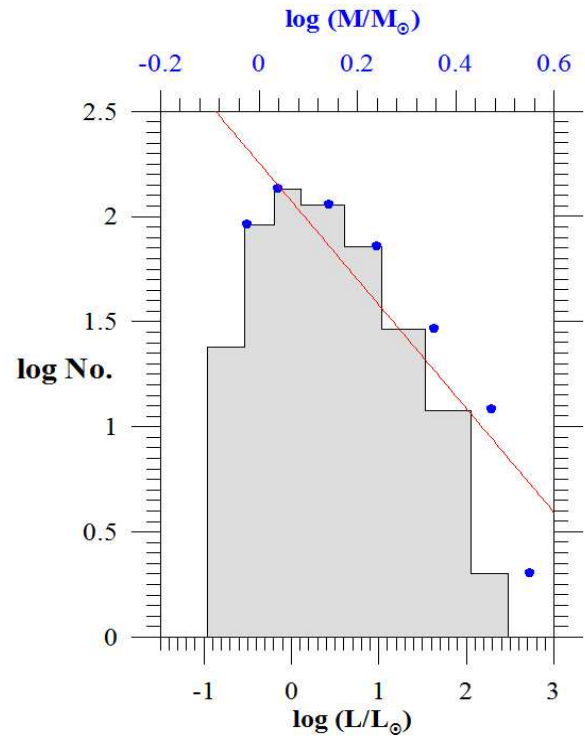}
\caption{The luminosity and mass functions of Dolidze 41. The gray histogram represents the luminosity distribution of the cluster, where the total luminosity is found to be -4.3 mag. The blue points refer to the mass distribution of the cluster, and the red line shows the linear fitting, where the initial mass function {\it IMF}-slope is found to be -2.3 $\pm$ 0.24.}
 \label{Fig-7}
\end{figure}

The stars have been counted in terms of the absolute magnitude $M_{G}$ after applying the distance modulus derived above. The magnitude bin interval are taken to be $\Delta M_{G}=0.50$\,mag. The magnitude bin intervals are elect to incorporate an affordable number of stars in every bin for the most effective potential statistics of the LF and MF. To convert the star absolute magnitude to luminosity and mass, we used the theoretical tables of evolutionary tracks of Marigo et al. \cite{2017ApJ...835...77M} of the same cluster's age. From the {\it LF} of Dolidze 41, we can infer that more massive stars are more centrally concentrated whereas the peak value lies at fainter magnitude bin of G $\approx$ 18.0 mag, i.e. M$_{G}$ $\approx$ 4.7 mag. The luminosity function has been created as shown in Fig.~\ref{Fig-7} (the gray histogram). In this context, the entire luminosity of the cluster is found to be $\sim$ --4.3 mag. The LF and MF are correlative to each other according to the well-known mass-luminosity relation. Therefore, the mass function distribution has been created as shown in Fig.~\ref{Fig-7} (the blue dotes), whereas the red line refers the initial mass function {\em IMF} slope, which can be obtained from the subsequent equation:
\begin{equation}
\frac{dN}{dM} \propto M^{-\alpha}
\end{equation}
where $\frac{dN}{dM}$ is the number of stars in the mass interval [M:(M+dM)] and $\alpha$ is the slope of the relation ($\Gamma$ = -2.3$\pm$0.24), which indicates that the estimated masses of Dolidze 41 lie in the range of 0.08 $M_{\odot}$ $<$ m$ <$ 0.5 $M_{\odot}$ according to Salpeter \cite{1955ApJ...121..161S}. We calculated the overall mass of the cluster by integrating the masses of the members \cite{2006AJ....132.1669S}, which is found to be 640 $M_{\odot}$.\\
\\
Knowing the overall mass of the cluster, and applying the equation of \cite{2001A&A...375..863J}, the tidal radius can be given as:
\begin{equation}
R_{t} = 1.46 (M_{c})^{1/3} ,
\end{equation}
where $R_{t}$ and $M_{c}$ are the tidal radius and the overall mass of the cluster respectively. $R_{t}$ is calculated to be 12.6 pc.
\\
\\
Following \cite{1971ApJ...164..399S}, the dynamical repose time {\em T$_R$} is obtained from the relation:
\begin{equation}
\large T_{R} = \frac{8.9 \times 10^{5} \sqrt{N} \times R_{h}^{1.5}}{\sqrt{m} \times log (0.4 N)}
\end{equation}
where $N$ is the number of the cluster members, $R_{h}$ is the radius containing half of the cluster mass in parsecs, and $m$ is the average mass of the cluster in solar unit, assuming that $R_{h}$ equals half of the cluster radius. Then, the repose time is found to be less than 40 Myr. The dynamical evolution parameter {\large $\tau$} = $Age/T_{R}$ $\approx$ 6.0, which means that the cluster Dolidze 41 is indeed dynamically relaxed.

\section{Conclusions}

The open cluster Dolidze 41 is poorly studied objet, the only photometric study, which found in the literature was carried out by TN \cite{2010arXiv1011.2934T} using 2MASS database. According our analysis for refining the fundamental parameters of Dolidze 41 in the {\em GAIA} era, we presented a real astro-photometric study here, which is somewhat different from the previous TN \cite{2010arXiv1011.2934T} one. The present and previous results are summarized and compared in table 1.

\begin{table}
\caption{Comparisons between the previous and present results.}
\begin{tabular}{lll}
\hline\noalign{\smallskip}Parameter& TN (2010)& Present work
\\\hline\noalign{\smallskip}
pm $\alpha$ cos $\delta$ &--~-- & -2.95 $\pm$ 0.14 mas/yr.\\
pm $\delta$ &--~-- & -4.68 $\pm$ 0.14 mas/yr.\\
Parallax & --~--& 0.23 $\pm$ 0.06 mas\\
Age& 400 Myr.& 200 Myr.\\
Metal abundance& 0.019&0.0152\\
$E(B-V)$& 0.53 mag.& 0.54 mag.\\
$R_{v}$& 3.25& 3.1\\
Intrinsic Modulus& 12.20 mag.& 13.30 $\pm$ 0.10 mag.\\
Distance& 1763 pc.& 4625 $\pm$ 210 pc.\\
Limited radius& 5.0$^{'}$ & 6.6$^{'}$ (8.8 pc.)\\
Core radius& --~--& 1.8$^{'}$ (2.4 pc)\\
Tidal radius& --~--& 12.6 pc.\\
Membership& --~--& 480 stars\\
$R_g$& 8.2 kpc.&  8.5 kpc.\\
X$_{\odot}$& --435 pc.& 1140 pc.\\
Y$_{\odot}$& 1708 pc.& 4480 pc.\\
Z$_{\odot}$& 31 pc.& 80 pc.\\
Total luminosity& --~--& --4.3 mag. \\
{\it IMF} slope& --~--&$\Gamma = -2.3 \pm 0.24$\\
Total mass&--~--& $\approx$ 640 $\mathcal{M}_{\odot}$ {\it (minimum)}\\
Repose time& --~--& $<$ 40 Myr.\\
$c$ & --~--&  $\approx$ 0.6 \\
$\tau$ & --~--&  $\approx$ 6.0 \\
\hline
\end{tabular}
\end{table}

\begin{acknowledgements}
This work has made use of data from the European Space Agency (ESA) mission Gaia processed by the Gaia Data Processing and Analysis Consortium (DPAC), $(https://www.cosmos.esa.int/web/gaia/dpac/consortium)$. Funding for the DPAC has been provided by national institutions, in particular the institutions participating in the Gaia Multilateral Agreement. In addition, the present study makes use of data products from the Two Micron All Sky Survey, which is a joint project of the University of Massachusetts and the Infrared Processing and Analysis Center/California Institute of Technology, funded by the National Aeronautics and Space Administration and the National Science Foundation. Virtual observatory tools like Topcat and Aladin have been used in the analysis.
\end{acknowledgements}


\end{document}